\documentclass[a4paper, 10pt, final]{amsart}

\usepackage[english]{babel}        
\usepackage[normalem]{ulem}        
\usepackage{amssymb, amsmath, ulem, amsthm}
\usepackage{enumerate}
\usepackage[small,bf]{caption}
\usepackage{graphicx,subfigure}

\newcommand{\eps}{\varepsilon}

\newcommand{\D}[2]{\frac{\partial{#1}}{\partial{#2}}}

\newtheoremstyle{dotless}{}{}{\itshape}{}{\bfseries}{}{ }{}

\newtheoremstyle{def}
	{0.5cm}                   
  {0.5cm}                   
  {}           							
  {}                      
  {\bfseries}  					  
  {}                      
  {\newline}        			
  {\underline{\thmname{#1} \thmnumber{#2:}} \thmnote{[#3]}}
  {}                       

\newcommand{\ul}{\underline}
\newcommand{\cal}{\mathcal}
\newcommand{\q}{\quad}
\newcommand{\tx}{\text}
\newcommand{\eb}{\textbf}

\newcommand{\ds}{\displaystyle}

\newcommand{\refeq}[1]{\text{eq.~}(\ref{#1})}
\newcommand{\reffig}[1]{\text{fig.~}\ref{#1}}

\newcommand{\nach}[1]{\stackrel{#1}{=}}

\begin{document}

\title[Application of Generalized Fokker-Planck Theory]{Application of Generalized Fokker-Planck Theory To Electron And Photon Transport In Tissue}
\author{Edgar Olbrant} 
\author{Martin Frank}
\address{RWTH Aachen University, Department of Mathematics \& Center for Computational Engineering Science, Schinkelstrasse 2, 52062 Aachen, Germany}
\email{olbrant@mathcces.rwth-aachen.de}
\email{frank@mathcces.rwth-aachen.de}

%
%
%
%
\begin{abstract}
We study a deterministic method for particle transport in tissue in selected medical applications. Generalized Fokker-Planck (GFP) theory \cite{LeaLar01} has been developed to improve the Fokker-Planck (FP) equation in cases where scattering is forward-peaked and where there is a sufficient amount of large-angle scattering. We compare grid-based numerical solutions to Fokker-Planck and Generalized Fokker-Planck (GFP) in realistic applications. Electron dose calculations in heterogeneous parts of the human body are performed. Accurate electron scattering cross sections are therefore included and their incorporation in our model is extensively described. Moreover, we solve GFP approximations of the radiative transport equation to investigate reflectance and transmittance of light in tissue. All results are compared with either Monte Carlo or discrete-ordinates transport solutions.
\end{abstract}

\maketitle

%
%
%
\section{Introduction}
\noindent A difficult and important challenge in electron and photon transport is still the numerical solution of the Boltzmann transport equation (BTE) \cite{Gar04, Bor98}. To contribute to this field of research, we study selected approximations of the transport equation for electrons and photons and present numerical results in different geometries. 

Nowadays cancer patients often undergo therapies with high energy ionizing radiation. In external radiotherapy photon beams dominate in clinical use whereas less patients receive electron therapy. Treatments are also performed by using heavy charged particles like ions or protons. These have higher costs for their particle accelerators but gain more and more importance due to first high intensity laser systems for protons \cite{SchPfoJae06}.

To aid the recovery of patients it is important to deposit a sufficient amount of energy in the tumor. Simultaneously, the ambient healthy tissue should not be damaged. Therefore, the success of such radiation treatments strongly depends on the correct dose distribution. It is recommended that uncertainties in dose distributions should be less than 2\% to get an overall desired accuracy of 3\% in the delivered dose to a volume \cite{AAPMreport85}. Additionally, thresholds have been developed to compare dose results computed by different algorithms. A suggested tolerance in homogeneous geometries is the 2\% (relative pointwise difference) or 2 mm (absolute distance to agreement) criterion. However, in heterogeneities this limit increases to 3\% or 3 mm \cite{VenWelMij01}.

Up to now many clinical dose calculation algorithms rely on pencil beam models. Originally developed for cosmic ray showers, Fermi (\cite{Fer40}, cited by \cite{RosGre41}) and Eyges \cite{Eyg48} introduced a small-angle scattering theory which was afterwards applied to electrons. This theory was used by Hogstrom et al.\ to propose the pencil-beam model. Their algorithm includes experimental data, taken from dose measurements in a water phantom, to compute the central-axis dose \cite{HogMilAlm81}. Although it was a first clinically applicable model its accuracy deep in the irradiated material or in heterogeneities is poor. This is basically due to the small-angle approximation in Fermi-Eyges theory, $\theta \approx \tan(\theta)$. It is true that single electron collisions show small deviations. However, after multiple scattering events they accumulate to big angle changes in large penetration depths. Besides crude approximations like small deviations of particles throughout their whole path through the tissue \cite{LarMifFra97}, geometric structures transverse to the beam direction are assumed to be infinite. Although many improvements of Fermi-Eyges theory were performed, e.g., by including additional correction factors \cite{Jet88, JetBie89, AhnSaxTre92, ShiHog91} they still suffer from the small anlge and the homogeneity assumption. Comparisons with experimental data showed disadvantages in inhomogeneous phantoms \cite{MarReyWag02}.

A statistical simulation method for radiation transport problems is the Monte Carlo method \cite{And91}. It performs direct simulations of individual particle tracks which result from a random sequence of free flights and interaction events. In this way random histories are generated. If their number is large enough macroscopic quantities can be obtained by averging over the simulated histories. Monte Carlo tools model physical processes very precisely and can handle arbitrary geometries without losing accuracy. Although they rank among the most accurate methods for predicting absorbed dose distributions, their high computation times limit their use in clinics. Due to the increase in computing power and decrease in hardware costs Monte Carlo techniques have recently become a growing field in radiotherapy \cite{SpeLew08}. Not only general-purpose Monte Carlo codes are now publicly available but also commercial Monte Carlo treatment planning systems \cite{SiaWalDAl01, CygLocDas05}. However, they have not yet gained widespread clinical use.

A different approach in the solution of radiation transport problems are deterministic calculations solving the linear BTE. In principle, its solution will give very accurate dose distributions comparable to Monte Carlo simulations. The BTE can be analytically solved only in very simplified geometries, which is insufficient for clinical applications. Additionally, numerical solutions to the BTE require deterministic methods coping with a six-dimensional phase space. Because of this and their simpler implementation, Monte Carlo techniques have so far prevailed in the medical physics community. Nevertheless, B\"orgers \cite{Bor98} argued that on certain accuracy conditions deterministic methods could compete with Monte Carlo calculations.

In this paper, we describe electron and photon transport in media by solving the Generalized Fokker-Planck (GFP) approximation of the linear BTE \cite{LeaLar01}. For electron transport, this is an extension of the Fokker-Planck (FP) model, formulated in \cite{HenIzaSie06}, to higher-order approximations in angle scattering. It should be stressed that this approach brings along several advantages: The BTE does not require any assumption on the geometry so that arbitrary heterogeneities are possible. Furthermore, we benefit from mathematical and physical approximation ans\"atze because they can be directly included in the differential equation. This avoids heuristic assumptions. In contrast to Monte Carlo simulation, deterministic solutions do not suffer from statistical noise and their resolution does not depend on the number of particles traversing a certain region. Moreover, the treatment planning problem can be formulated as a PDE-constrained optimization problem. This structure can be used to obtain additional information to speed-up the optimization \cite{FraHerSan09, FraHerSch08}. This is hard to achieve by Monte Carlo techniques.

Previous studies in deterministic methods for radiotherapy primarily concentrated on a combination of rigorous analytic solutions and laboratory measurements (see pencil beam models above). Without explicitly using experimental data in their model, Huizenga and Storchi \cite{HuiSto89} presented the phase space evolution (PSE) model for electrons and subsequently applied it to multilayered geometries \cite{MorHui92}. Various improvements and extension to 3D beam dose calculations were performed \cite{JanRieMor94, JanKorSto97}. However, 3D dose calculations showed disadvantages in computation times \cite{KorAkhHei00}. Moreover, PSE models use first-order discretizations in space which cannot compete with MC techniques \cite{Bor98}. By contrast, our access focuses on the continuous model of the Boltzmann equation discretized with high-order schemes.

First studies for deterministic dose calculations in charged particle transport came from well-known procedures for numerical solutions to the transport equation of neutrons or photons. Several approximative models to the BTE have been developed. Each of these methods has its advantages and drawbacks \cite{Bru02}. Multi-group methods are sometimes used to discretize the energy domain \cite{DatAltRay96}. This leads to a number of monoenergetic equations to be separately discretized in space and angle. Whereas discretizations in space are usually done by finite difference and finite element (FE) methods, the remaining angle domain is discretized by discrete ordinates \cite{Bal00, Eds05}. Such an approach is implemented in the solver Attila \cite{GifHorWar06}. Recently, 3D dose calculations for real clinical test cases were performed with Attila \cite{VasWarDav08}. Results with similar accuracy as Monte Carlo calculations were achieved in promising computation times.

Less frequent are angular FE approaches \cite{CopRav95} seeking to reduce ray effects. Tervo et al.\ extended FE discretizations to all variables in spatial, angular and energy domains \cite{TerKolVau99} so that no group cross sections were needed. Moreover, a detailed description of three coupled BTEs with FE discretizations of all variables was proposed in \cite{BomTerVau05}.

In this paper, the BTE is approximated by the continuous slowing down (CSD) method \cite{LarMifFra97}. Scattering processes are modelled by Generalized Fokker-Planck approximations \cite{LeaLar01} and compared with classical Fokker-Planck \cite{Pom92} solutions. Leakeas and Larsen \cite{LeaLar01} showed that scattering kernels with a sufficient amount of large-angle scattering yield inaccurate FP results. As an extension of the work in \cite{LeaLar01}, we perform deterministic GFP simulations and investigate their behaviour in real applications. It is well-known that electron scattering is dominantly forward-peaked. Hence, many electron transport simulations used the classical FP approximation \cite{HenIzaSie06, FraHenKla07, DucMorTik09}. However, up to now no comparisons with GFP dose computations have been done including realistic physical scattering cross sections. In our case the latter are extracted from ICRU libraries \cite{ICRU07}. We describe in detail how transport coefficients in the BTE can be computed from these scattering cross sections and compute GFP electron dose profiles in inhomogeneous geometries.

Even more challenging for GFP theory are scattering kernels including large angle-scattering. We therefore investigate transport of photons in tissues with forward-peaked and large-angle scattering. Using test cases from \cite{RodKim08} the radiative transport equation for GFP approximations up to order five is solved to determine reflectance and transmittance of light in tissue.

In the remaining paper you will find the following structure: A short discussion of basic electron interactions with matter is given at the beginning. We describe a model for electron transport and review crucial steps of the GFP theory. In addition, we derive transport coefficients for the GFP equations from databases for electron scattering cross sections. In Section 3, a deterministic model for light propagation together with different scattering kernels is introduced. Discretization methods used in our GFP equations are studied in Section 4. In Section 5, we compute FP, GFP and discrete-ordinates results for transmittance and reflectance of light by a slab. Moreover, we numerically compare FP, GFP and Monte Carlo solutions for 5 and 10 MeV electrons in homogeneous and heterogenous slab geometries. Section 6 gives the conclusions and outlooks. Appendix A contains explicit formulae for high-order polynomial operators. In Appendix B, we present the equations to be solved for the GFP coefficients.
%
%
%
%
%
%
%
%
\section{Deterministic Model for Electron Transport}
\subsection{Physical Interactions}
Electron beams are nowadays a widely spread tool in cancer therapy. Typical electron beams, provided by high energy linear accelerators, range from 1 to 25 MeV. During irradiation of human tissue electrons interact with matter through several competing mechanisms:
\begin{enumerate}[1.]
\item \textit{Elastic Scattering:} This is usually a non-radiative interaction between electrons and the atomic shell. Projectiles experience small deflections and lose little energy. High energy electrons can also penetrate through atomic shells and are afterwards scattered at the bare nucleus without any energy loss. With kinetic energies above 1 keV elastic scattering in water dominantly occurs in the forward direction \cite{LaVPim97}.

\item \textit{Soft Inelastic e$^-$-e$^-$ Scattering:} Electrons interact with other electrons of the outer atomic shell which usually leads to excitation or ionisation of the target particle. Here binding energies are only a few eV so that projectile electrons transfer little energy and are hardly deflected.

\item \textit{Hard Inelastic e$^-$-e$^-$ Scattering:} These collisions are determined by large transfer energies to the target electron. What 'large' exactly means is specified in MC codes by cutoff energies. In PENELOPE \cite{Pen09}, for example, the default value of this simulation parameter is set to 1 \% of the maximum energy of all particles. As a consequence, the target electrons are ejected with larger scattering angles and higher kinetic energies (delta rays). They act as an additional source in the transport equation.

\item \textit{Bremsstrahlung:} Caused by the electrostatic field of atoms, electrons are accelerated and hence emit bremsstrahlung photons. However, for energies below 1 MeV this phenomenon can be neglected. Bremsstrahlung photons are not mainly emitted in the forward direction. The lower their kinetic energy the more isotropic their angle distribution becomes \cite{Kri07}.
\end{enumerate}
Evidently, there are more interaction processes like ejection of Auger electrons or characteristic X-ray photons. But they are very unlikely in the energy range considered.

Although inelastic collisions are decisive for the energy transfer, the radiation damage in the patient strongly depends on the spatial distribution of electrons in their passage through matter. They dominantly undergo multiple scattering events with small deviations. However, single backward scattering events also occur frequently which leads to tortuous trajectories of electrons. To a big extent such trajectories are due to elastic collisions. We therefore focus on very accurate and realistic simulation of elastic processes in our model. This is achieved by transport coefficients extracted from the ICRU 77 database \cite{ICRU07}. Inelastic transport coefficients are obtained in the same way.

Elastic and soft inelastic events lead to small energy loss. With kinetic energies above 1 keV, electrons are assumed to lose their energy continuously \cite{Gar05}. Because of this, we implement the CSD approximation to model energy loss of electrons. Hence, we neglect large energy loss fluctuations caused by hard inelastic collisions.

Bremsstrahlung effects are considered in our model in a very restricted way: Only energy transfer of electrons after soft bremsstrahlung collisions are simulated by means of the radiative stopping power. However, the effect of hard photon emission as well as the transport of photons are disregarded so far.
\subsection{The Generalized Fokker-Planck Approximation for Electrons} \label{GFPforElectrons}
\noindent Particles can be described by a six dimensional phase space $(\uline{r}, E, \uline{\Omega}) \in (V\times I \times S^2)$ with
\begin{align*}
\text{\uline{spatial variable}} \quad	& \uline{r}=(x_1,x_2,x_3)\in V \subset \mathbb{R}^3 \; \text{open and bounded}  \\
\text{\uline{energy variable}} \quad	& E \in I=[E_{\tx{min}}, E_{\tx{max}}] \subset \mathbb{R}  \\
\text{\uline{direction variable}} \quad & \uline{\Omega}=(\Omega_1,\Omega_2,\Omega_3) \in S^2 \subset \mathbb{R}^3 \; \text{unit sphere} \\
	 &\, \, \, \, \, =(\sqrt{1-\mu^2}\cos(\phi),\sqrt{1-\mu^2}\sin(\phi),\mu), \q \mu=\cos(\theta).
\end{align*}
If particle transport takes place in an isotropic and homogeneous medium in which interaction processes are Markovian and particles do not interact with themselves, their distribution can be described by the unique solution of the time-dependent linear BTE. For practical applications in radiotherapy we are faced with issues like distributions of the deposited energy in tissue or penetration depths of the beam. For many purposes it is therefore sufficient to know the steady solution:
\begin{equation}
\sigma_a \Psi(\ul{r}, E, \ul{\Omega}) + \ul{\Omega}\cdot\nabla\Psi(\ul{r}, E, \ul{\Omega}) = L_B \Psi(\ul{r}, E,\ul{\Omega}) \label{MBE}
\end{equation}
$\text{\q with \q} \q  \Psi(\uline{r}, E, \uline{\Omega})=\Psi_b(\uline{r}, E,\uline{\Omega}) \q \q \forall \uline{r} \in \partial V,\; \uline{\Omega} \cdot \uline{n}<0, \; E\in I.$ \\ \\
$\Psi(\ul{r}, E, \ul{\Omega})$ is called {\it angular flux} and denotes the distribution of particles travelling in direction $\ul{\Omega}$. Boundary conditions (BC) are imposed on the angular flux depending on the incident beam. Absorption of particles is described by the {\it absorption cross section} $\sigma_a$. The right hand side
\[
L_B \Psi(\ul{r},E, \ul{\Omega}):= \int_0^{\infty} \int_{4\pi} \sigma_{s}(E',E,\ul{\Omega}\cdot\ul{\Omega}') \Psi(\ul{r}, E',\ul{\Omega}') d\Omega' dE' - \Sigma_s(E) \Psi(\ul{r},E, \ul{\Omega})
\]
is known as linear {\it Boltzmann Operator} and describes scattering. Its integral contains the {\it differential scattering cross section} (DSCS) $\sigma_s(E',E,\ul\Omega\cdot\ul\Omega')$ characterising interaction mechanisms in which particles are deflected. The dot product $\ul{\Omega} \cdot \ul{\Omega}'=\cos(\theta_0)=\mu_0$ indicates that the scattering probability only depends on the scattering angle. This implies that the deflection of scattered particles is axially symmetrical around the direction of incidence $\ul{\Omega}'$. Integrating the scattering kernel $\sigma_{s}(E',E,\ul{\Omega}\cdot\ul{\Omega}')$ over all angles and energies, one gets the {\it total differential scattering cross section} (TDSCS)
\begin{equation}
	\Sigma_s(E) = 2\pi \int_0^{\infty} \int_{-1}^1 \sigma_s(E', E,\mu_0)d\mu_0 dE'. \label{tdscs2} 
\end{equation}%
The angle and energy integral in the Boltzmann operator is the main difficulty in its numerical solution. That is why an important aim is to develop accurate approximations. However, up to now there is no predominant method used for all types of particles. In fact, depending on specific particle properties, one has to choose an appropriate approximation.

One crucial property of elastic DSCSs in water is a sharp peak in the forward direction \cite{ItiMas05}. To express this mathematically we define the positive {\it n-th scattering transport coefficient} (STC) 
\begin{align}
\xi_n(E):=2\pi \int_0^{\infty} \int_{-1}^{1}(1-\mu_0)^n \sigma_s(E',E,\mu_0)d\mu_0 dE' \quad \tx{for all } n \geq 0
\end{align}
and assume that as n increases, the coefficients $\xi_n$ fall off sufficiently fast, i.e.,
\begin{align}
\xi_{n+1}(E) \ll \xi_n(E) \quad \tx{for all } n \geq 0 \tx{ and } E \in I. \label{fpeak}
\end{align}
Additionally, elastic scattering often entails small energy loss. A first approximation is therefore to expand the scattering kernel around $\mu_0=1$ and $E=E'$. In this way Pomraning \cite{Pom92} showed that the already known FP operator is the lowest-order asymptotic limit of the integral operator $L_B$. In \cite{LeaLar01} this FP operator is derived as a first order angle approximation to $L_B$: \\
\uline{Fokker-Planck} \hspace{2cm}$L_{FP}:=\frac{\xi_1}{2}L$
\begin{align}
\tx{where} \q \q \q L_B\Psi(\uline{\Omega})=L_{FP}\Psi(\uline{\Omega})+{\cal O}(\eps) \q \tx{for } \eps \ll 1.  \label{LB_FP}
\end{align}
Since the spherical Laplace-Beltrami operator
\begin{align*}
L = \left[ \frac{\partial}{\partial \mu}(1-\mu^2)\frac{\partial}{\partial\mu} + \frac{1}{1-\mu^2}\frac{\partial^2}{\partial\phi^2} \right] \quad \text{with} \; \mu=\cos(\theta)
\end{align*}
is differential in angle the nonlocal \textit{integral} Boltzmann operator $L_B$ is now approximated by a local \textit{differential} operator. The crucial point is that an integro-differential equation is transformed into a partial differential equation. Although discretizations of differential equations often lead to large linear systems their numerical effort turns out to be much lower. This is due to the local character of differential equations which bring along much sparser matrices. \\
Pomraning's resulting Fokker-Planck equation for particle transport in an isotropic medium yields:
\begin{align}
\sigma_a \Psi(\ul{r},E,\ul{\Omega}) + \ul{\Omega}\cdot\nabla\Psi(\ul{r},E,\ul{\Omega}) &=  \frac{\xi_1(E)}{2} L \Psi(\ul{r},E,\ul{\Omega}) + \frac{\partial (S(\ul{r},E) \Psi(\ul{r},E,\ul{\Omega}))}{\partial E}.
\end{align}
$S(\ul{r},\eps)$ is called stopping power defined by
\begin{align}
S(\ul{r},E) = \int_{0}^{\infty} \int_{4\pi} (E -E') \sigma_s(E',E,\ul{\Omega}\cdot \ul{\Omega}')  \, d\Omega dE'. \label{StpPom}
\end{align}
The standard Fokker-Planck approximation is a frequently used method to describe transport processes in media with \textit{highly} forward-peaked scattering. Comparisons with real data, however, reveal that many scattering processes of interest contain a small but sufficient amount of large-angle scattering. To gain higher order asymptotic approximations to $L_B$ one could expand $L_{B}$ to a \\ \\
\uline{Polynomial Operator} \quad $L_{Pn}:=\sum_{m=1}^{n}a_{n,m}L^m$ with $a_{n,m} \in {\cal O}(\xi_m)={\cal O}(\eps^{m-1})$
	\begin{align*}
		\tx{with} \q L_B\Psi(\uline{\Omega})=L_{Pn}\Psi(\uline{\Omega})+{\cal O}(\eps^n) \q \tx{for all } n\geq 1.
	\end{align*}
However, Leakeas and Larsen showed that eigenvalues of $L_{Pn}$ might become positive so that the angular flux could become infinite \cite{LeaLar01}. This served as motivation for them to develop asymptotically equivalent operators to $L_{Pn}$ which remain stable and preserve certain eigenvalues of $L_B$. We have summarized the details of the computation in the appendix. We end up with Generalized Fokker-Planck (GFP) equations which incorporate large-angle scattering and are therefore more accurate than the conventional Fokker-Planck equation:
\begin{align}
\sigma_a \Psi(\ul{r},E,\ul{\Omega}) + \ul{\Omega}\cdot\nabla\Psi(\ul{r},E,\ul{\Omega}) &= L_{GFP_n} \Psi(\ul{r},E,\ul{\Omega}) + \frac{\partial (S(\ul{r},E) \Psi(\ul{r},E,\ul{\Omega}))}{\partial E}.
\end{align}
For positive coefficients $\alpha_i(E), \beta_i(E)$ and $m\in \mathbb{N}$, the GFP operators are defined by 
\begin{align*}
L_{GFP_{2m}} := \sum_{i=1}^m \alpha_i(E) L(I-\beta_i(E) L)^{-1} \quad \text{ and } \quad L_{GFP_{2m+1}} := L_{GFP_{2m}} +\alpha_{m+1}(E)L.
\end{align*}
%
%
%
%
\subsection{Determination of Physical Quantities}
The fundamental part in GFP theory is based on the assumption of forward-peakedness (\refeq{fpeak}). Its expected accuracy strongly depends on the behaviour of transport coefficients $\xi_n(E)$ which differ for different projectiles and materials. The ICRU Report 77 provides differential cross sections for elastic and inelastic scattering of electrons and positrons for different materials and energies between 50 eV and 100 MeV \cite{ICRU07}. To obtain transport coefficients $\xi_n(E)$ we use these cross sections and proceed in the following way:
\begin{enumerate}[(a)]
\item {The ELSEPA code system, distributed with the report, calculates elastic and inelastic angular differential cross sections for a fixed energy $E$
\begin{align*}
\sigma^{\tx{el,inel}}(E,\mu_0)=\int_0^{\infty} \sigma^{\tx{el,inel}}_s(E',E,\mu_0) dE'
\end{align*}
in tabulated form for discrete $\mu_0$ and $\sigma^{\tx{el,inel}}(E,\mu_0)$. For a predetermined set of energy values between 50eV and 100MeV, data for $\sigma^{\tx{el,inel}}(E,\mu_0)$ are extracted from these files.
}
\item {With that we calculate the $n$-th transport coefficient for a fixed energy $E$
\begin{align*}
\xi_n^{\tx{el,inel}}(E)=2\pi \mathcal{N} \int_{-1}^1 (1-\mu_0)^n \sigma^{\tx{el,inel}}(E,\mu_0) d\mu_0 \q \tx{with} \q \mu_0=\cos(\theta_0)
\end{align*}
via numerical integration of the tabulated cross sections $\sigma^{\tx{el,inel}}(E,\mu_0)$ by means of the trapezodial rule. Additionally, we multiply the result by the molecular density of the transmitted matter 
\begin{align*}
\mathcal{N}=N_A\frac{\rho}{A}.
\end{align*}
$N_A$ is Avogadro's number, $\rho$ the mass density of the material and $A$ its molar mass. 
}
\item Again, all computed results of $\xi_n^{\tx{el,inel}}(E)$ are stored and used as a look-up table. To obtain the $n$-th transport coefficient at the desired energy $E$ this tabulated data is linear interpolated.
\end{enumerate}
Finally, we use the following transport coefficient in our equation:
\begin{align*}
\xi_n(E) := 2\pi\mathcal{N} \int_{-1}^1 (1-\mu_0)^n (\sigma^{\tx{el}}(E,\mu_0) +\sigma^{\tx{inel}}(E,\mu_0)) d\mu_0.
\end{align*}
\begin{figure}%
\centering \includegraphics[scale=0.65]{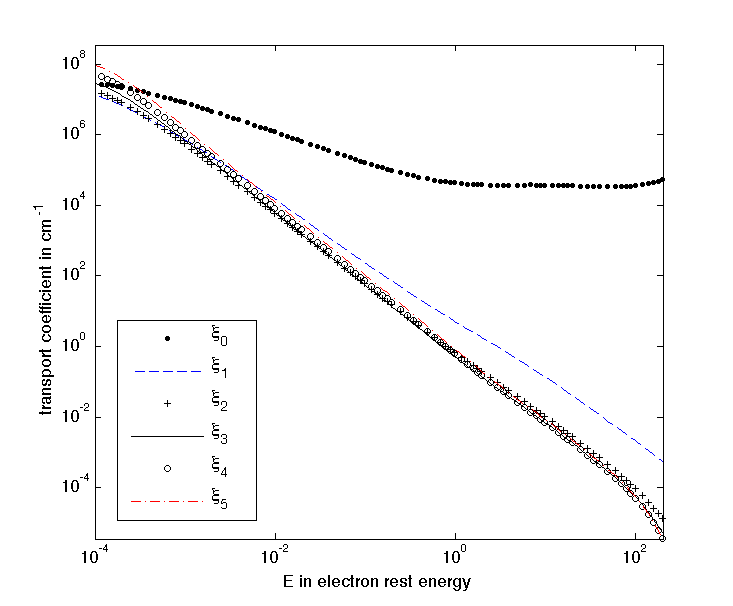}%
\setcaptionwidth{.9\textwidth}
\caption{Comparison of scattering transport coefficients in liquid water for energies between 50eV and 100MeV.} \label{STCtot}%
\end{figure}%
Fig.~\ref{STCtot} illustrates electron transport coefficients of different order in liquid water. Except for the $0$-th, every coefficient is strictly monotonically decreasing. For $E\geq 10^{-3}$, $\xi_1$ is always bigger than $\xi_2$ but, as $E$ decreases, their deviation reduces more and more. Unfortunately, for increasing $n\geq 2$ the difference between two consecutive $\xi_n$ is so small that the assumption of forward-peakedness is not fulfilled. The inelastic transport coefficients are much smaller than elastic transport coefficients for $n \geq 1$. The higher the order of the transport coefficient the larger is the difference between elastic and inelastic ones.

\noindent Our stopping power in \refeq{StpPom} is equivalent to the physical total stopping power as the sum of collision and radiative stopping powers. For different materials both are directly included in the files of the ICRU database. Hence, we do a linear interpolation to get the stopping power $S(E)$ at the desired energy $E$.
%
%
%
%
%
%
%
%
\section{Deterministic Model for Light Propagation}
\subsection{The Generalized Fokker-Planck Equation for Grey Photons}
Many medical applications like cancer treatment or optical imaging of tumors make use of propagation of laser light in tissue. Its behaviour is determined by the solution of the steady grey radiative transport equation
\begin{align}
\sigma_a \Psi(\ul{r},\ul{\Omega}) + \ul{\Omega}\cdot\nabla\Psi(\ul{r}, \ul{\Omega}) = \mu_s \left[ \int_{4\pi} \sigma_s(\ul{\Omega}\cdot\ul{\Omega}')\Psi(\ul{r},\ul{\Omega}) d\Omega' -\Psi(\ul{r},\ul{\Omega}) \right] \label{RTE}
\end{align}
$\text{\q with \q} \q \q \Psi(\uline{r},\uline{\Omega})=\Psi_b(\uline{r},\uline{\Omega}) \q \forall \uline{r} \in \partial V,\; \uline{\Omega} \cdot \uline{n}<0.$ \\

\noindent Similar to \ref{GFPforElectrons} one can derive GFP approximations to the radiative transport equation in a straightforward way. The intensity $\Psi(\uline{r},\uline{\Omega})$ describes the radiation power flowing in direction $\Omega$ which is influenced by scattering and absorption coefficients $\sigma_a$ and $\mu_s$. More important is the scattering kernel $\sigma_s(\ul{\Omega}\cdot\ul{\Omega}')$. It is also characteristic for biological tissue that this kernel has a sharp peak at $\ul{\Omega}\cdot\ul{\Omega}'=1$. For its simulation mathematically simple scattering kernels with a free parameter are used.
\subsection{Models for Scattering Kernels}
One often cited scattering kernel is the {\it (single) Henyey-Greenstein kernel} defined by
\begin{align}
\sigma_s^{HG}(\mu_0) &= \frac{\Sigma_s^{HG}}{2\pi} f_{HG}(\mu_0) \label{HGkernel}  \\
\text{where} \q \q \q f_{HG}(\mu_0) &:= \frac{1-g^2}{2(1 -2g\mu_0 +g^2)^{3/2}} \q \tx{for} \q g\in (-1;1) \nonumber
\end{align}
is the corresponding phase function and $\Sigma_s^{HG}$ the TDSCS. The single parameter $g$ determines the amount of small- and large-angle scattering. If $g\approx 1$, $\sigma_s^{HG}$ is not only strongly forward-peaked but also includes large-angle scattering. Its value depends on the irradiated tissue. For human tissue typical values for $g$ are around $0.9$ (human blood: $g=0.99$, human dermis: $g=0.81$ \cite{ChePraWel90}). Expanding $f_{HG}(\mu_0)$ in Legendre polynomials gives expressions for $\xi_n$ depending only on g:
\begin{align}
	\xi_1 &= 1-g \\
	\xi_2 &= \frac{4}{3} -2g +\frac{2}{3}g^2 \\
	\xi_3 &= 2 -\frac{18}{5}g +2g^2 -\frac{2}{5}g^3 \\
	\xi_4 &= \frac{16}{5} -\frac{32}{5}g +\frac{32}{7}g^2 -\frac{8}{5}g^3 +\frac{8}{35}g^4 \\
	\xi_5 &= \frac{16}{3} -\frac{80}{7}g +\frac{200}{21}g^2 -\frac{40}{9}g^3 +\frac{8}{7}g^4 -\frac{8}{63}g^5
\end{align}
This is all we need to calculate $\alpha_i$ and $\beta_i$ for GFP$_2$ to GFP$_5$. It turns out that they remain throughout positive.

\noindent To control large-angle and forward-peaked scattering a linear combination of forward and backward Henyey-Greenstein phase functions was introduced \cite{RodKim08}. For real constants $g_1 \in (-1;0], \, g_2 \in [0;1), \, b \in [0;1]$ and the phase function
\begin{align}
f_{DHG}(\mu_0) &:= b f_{HG}(\mu_0,g_1) +(1-b)f_{HG}(\mu_0,g_2),
\intertext{the {\it double Henyey-Greenstein} scattering kernel is defined by}
\sigma_s^{DHG}(\mu_0) &= \frac{\Sigma_s^{DHG}}{2\pi} f_{DHG}(\mu_0).   \label{DHGkernel}
\end{align}
An indicator for the amount of forward or backward scattering is the constant $b$: Setting $b=0$ the backward scattering phase function $f_{HG}(\mu_0,g_1)$ vanishes whereas $b=1$ reduces $f_{DHG}(\mu_0)$ to the single Henyey-Greenstein phase function $f_{HG}(\mu_0,g_1)$. This provides an opportunity to adapt it to the material of interest.

\noindent Analogous to above procedure for the single HG phase function one can conclude that
\[ \sigma_{sn} =b(g_1^n -g_2^n) +g_2^n \] 
from which all STCs $\xi_n$ can be calculated.
%
%
%
%
%
%
%
\section{Numerics for Generalized Fokker-Planck}
\subsection{Discretization of Differential GFP Equations} \label{MonDiffGFP}
Replacing the right-hand side of the Boltzmann equation by the GFP$_2$ approximation operator yields:
\begin{align}
S(\ul{r},E) \frac{\partial\Psi(\ul{r}, E,\ul{\Omega})}{\partial E} + \sigma_a \Psi(\ul{r},E,\ul{\Omega}) + \ul{\Omega}\cdot\nabla\Psi(\ul{r},E,\ul{\Omega}) &=  \alpha(E) L (I- \beta(E)L)^{-1} \Psi(\ul{r},E,\ul{\Omega}) \nonumber \\
&+ \Psi(\ul{r},E,\ul{\Omega})) \frac{\partial S(\ul{r},E)}{\partial E}.   \label{GFP2_MBE}
\end{align}
To solve \refeq{GFP2_MBE} we restate it by setting
\begin{align}
\Psi^{(0)} = \Psi(\ul{r},E,\ul{\Omega}) \q \tx{and} \q \Psi^{(1)} = (I-\beta(E) L)^{-1} \Psi^{(0)}, \nonumber
\end{align}
so that it becomes
\begin{align}
S(\ul{r},E) \frac{\partial\Psi^{(0)}(\ul{r}, E,\ul{\Omega})}{\partial E} + \sigma_a \Psi^{(0)}(\ul{r},E,\ul{\Omega}) + \ul{\Omega}\cdot\nabla\Psi^{(0)}(\ul{r},E,\ul{\Omega}) &=  \alpha(E) L \Psi^{(1)}(\ul{r},E,\ul{\Omega}) \nonumber \\
&+ \Psi^{(0)}(\ul{r},E,\ul{\Omega})) \frac{\partial S(\ul{r},E)}{\partial E} \\
(I-\beta(E) L)\Psi^{(1)}(\ul{r},E,\ul{\Omega}) &= \Psi^{(0)}(\ul{r},E,\ul{\Omega}).
\end{align}
These equations form a coupled system of second-order differential equations with the angular momentum operator L. Solving this system requires differencing schemes in space and angle. Initial and boundary conditions are imposed on $\Psi^{(0)}$.  \\
The asymptotic GFP analysis transformed the original BTE into a new type of equation which requires an additional condition to the energy variable:
\begin{align}
\Psi^{(0)}(\ul{r},E_{\infty},\ul{\Omega})=0. \label{IC}
\end{align}
$E_{\infty}$ denotes a large cutoff energy. In the numerical simulations, it should be bigger than the energy of all particles from the incoming beam. \\
\noindent A simplified model to be studied is that of a plate which is infinitely extended in x and y directions with a thickness d in z direction. Due to symmetry reasons
\begin{itemize}
	\item{the angular flux is independent of x and y directions and} 
	\item{its direction of motion $\ul{\Omega}$ only depends on $\theta$.}
\end{itemize}
That is why the initial six dimensional problem has now been reduced to a three dimensional one. Although it seems that we only describe a one dimensional object in space one should not forget the fact that our \textit{slab} still remains three dimensional. From the mathematical point of view the symmetry of this model, however, leads to a one dimensional problem in space which decreases computational costs.

\noindent In slab geometry the aforementioned system reduces to
\begin{align}
S(\ul{r},E)\frac{\partial\Psi^{(0)}(z,E,\mu)}{\partial E} &= \alpha(E) L_{\mu} \Psi^{(1)}(z,E,\mu) + \Psi^{(0)}(\ul{r},E,\ul{\Omega})) \frac{\partial S(\ul{r},E)}{\partial E} \nonumber \\ 
& - \sigma_a \Psi^{(0)}(z,E,\mu) - \D{\Psi^{(0)}(z,E,\mu)}{z}\cdot \mu \label{rhs1} \\
(I-\beta(E) L_{\mu}) & \Psi^{(1)}(z,E,\mu) = \Psi^{(0)}(z,E,\mu), \label{I-betaL_Psi1}
\intertext{where the one dimensional angular momentum operator $L_{\mu}$ is defined by}
L_{\mu} &:= \D{}{\mu} \left [(1-\mu^2) \D{}{\mu} \right].
\end{align}
We solve this system in two steps:
\begin{enumerate}
	\item{Obtain the solution $\Psi^{(1)}(z,E,\mu)$ to \refeq{I-betaL_Psi1}.}
	\item{Plug $\Psi^{(1)}(z,E,\mu)$ in \refeq{rhs1} and solve the resulting differential equation.}
\end{enumerate}
To achieve accurate results and lower computation times a high-order scheme was implemented \cite{Mor85}:
\begin{align}
\tilde{L}_{\mu} \Psi^{(1)}(\mu_j) &= \frac{1}{w_j} \left [  D_{j+1/2} \frac{\Psi^{(1)}_{j+1} -\Psi^{(1)}_{j}}{\mu_{j+1} -\mu_j} - D_{j-1/2} \frac{\Psi^{(1)}_{j} -\Psi^{(1)}_{j-1}}{\mu_{j} -\mu_{j-1} }   \right ]  \label{morel} \\
D_{j+1/2} &= D_{j-1/2} -2\mu_{j}w_{j} \q \tx{with} \q D_{1/2}=0=D_{M+1/2}, \nonumber
\end{align}
where $\mu_{j}$ are abscissas for a Gauss-Legendre quadrature rule with weights $w_{j}$.

With knowledge of the already calculated explicit values $\Psi^{(1)}_{i,j}$ the right-hand side of \refeq{rhs1} reduces to a single $\Psi^{(0)}$ dependence. The resulting partial differential equation is discretized with finite differences in the z-direction. Hence, we end up with an ordinary differential equation in the energy variable $E$. Its solution is obtained by the embedded 2nd/3rd order Runge-Kutta MATLAB solver {\tt ode23} solving from the initial condition in \refeq{IC} backward in energy to $E=0$. The remaining discretizations are of first order in $z$ and of second order in $\mu$. 

\noindent Higher order GFP equations are discretized and solved analogously. However, due to more frequent occurrence of $L_{\mu}$ and $(I-\beta(E) L_{\mu})$ the right-hand side of \refeq{rhs1} becomes more involved and more systems have to be solved.
%
%
%
%
%
%
%
\section{Numerical Results}
%
%
%
%
\subsection{Slab Geometry: HG Kernel} \label{slabHGSAM}
\begin{figure}
\centering \includegraphics[scale=0.64]{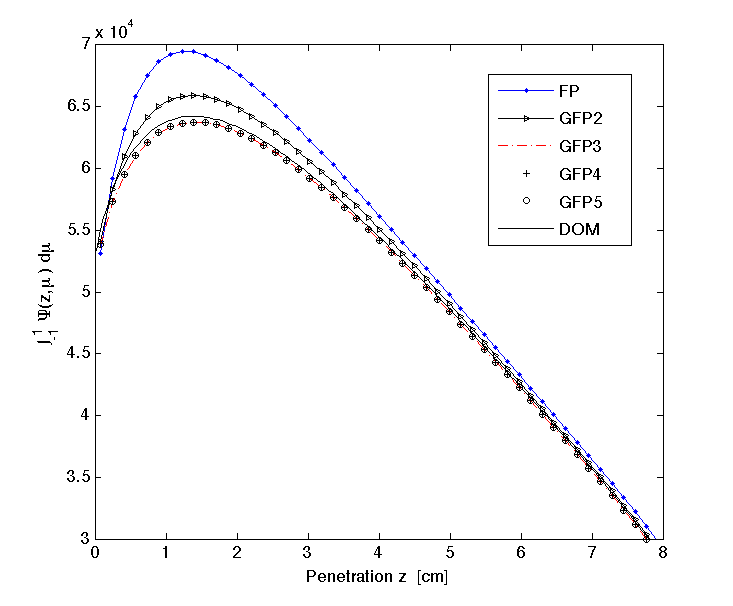} %
\caption{HG kernel in slab geometry with g=0.80. The difference between GFP$_3$-GFP$_5$ lines is so small that they almost agree here.} \label{HG08_1}%
\centering \includegraphics[scale=0.64]{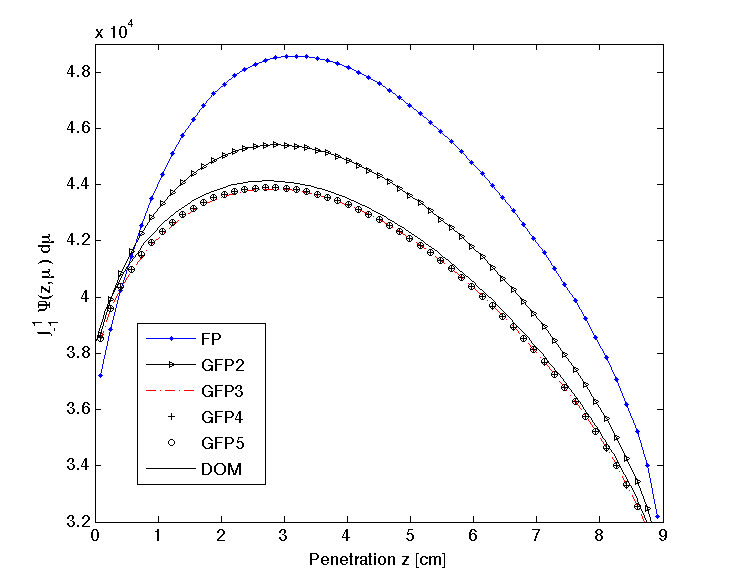} %
\caption{HG kernel in slab geometry with g=0.95. Solutions for GFP$_3$-GFP$_5$ already overlap.} \label{HG095_1}%
\end{figure}

First we neglect absorption and start with a simpler form of the GFP equation
\begin{align}
\ul{\Omega}\cdot\nabla\Psi(\ul{r},\ul{\Omega}) = L_{GFP_n} \Psi(\ul{r},\ul{\Omega}) 
\end{align}
This is solved in slab geometry with the HG scattering kernel from \refeq{HGkernel} for selected values of the anisotropy factor g. Symmetry properties mentioned above yield the following boundary value problem (exemplarily stated for GFP$_2$):
\begin{align}
\mu \D{\Psi^{(0)}(z,\mu)}{z} &= \alpha L_{\mu}\Psi^{(1)}(z,\mu) \label{ICBC_1} \\ \nonumber
(I-\beta L_{\mu})\Psi^{(1)}(z,\mu) &= \Psi^{(0)}(z,\mu)
\end{align}
\begin{align*}
	\ul{\text{BC}}:& \quad \Psi^{(0)}(0,\mu) =10^5 \cdot e^{-10(1-\mu)^2} \hspace{2.9cm}  1\geq \mu>0 \\ \nonumber
								 & \quad \Psi^{(0)}(d,\mu) =0  \hspace{4.7cm} -1\leq \mu<0
\end{align*}
For $g=0.8$ and $g=0.95$ results were computed by time marching with an adaptive Runge-Kutta solver until a steady state was reached. Incoming photons moving in positive z-direction at $z=0$ are simulated by narrow Gaussian peaks around $\mu=1$. Corresponding graphs illustrate steady solutions and use penetration depth in cm as x- and $\int_{-1}^{1}\Psi(z,\mu,s) d\mu$ in 1/(Jcm$^2$s) as y-axis. The latter quantity is sometimes called energy density and is related to the dose. Discretization parameters in z (110 points) and in $\mu$ (64 points) direction were chosen large enough to reach convergence. Figs.~\ref{HG08_1}-\ref{HG095_1} additionally show converged transport solutions generated by a discrete ordinates method (DOM) which we use as benchmark in the following. 

%
%
Each distribution forms a monotonically increasing function until it reaches a maximum and strictly decreases afterwards. There is a large difference between FP, GFP$_2$ and the other GFP approximations. However, GFP$_3$-GFP$_5$ are hardly distinguishable. For increasing values of g this discrepancy between GFP$_3$ and GFP$_5$ becomes bigger whereas results for GFP$_4$ and GFP$_5$ show throughout no distance at all. As to the DOM curve we observe that FP and GFP$_2$ give poor approximations for small penetration depths. However, GFP$_{3,4,5}$ give quite good solutions throughout the whole penetration range.
%
%
%
%
\subsection{Single Slab: Light Propagation in Tissue}
Gonz\'alez-Rodr\'iguez and Kim studied light propagation in tissue including both forward-peaked and large-angle scattering \cite{RodKim08}. They examined several approximation methods and especially implemented GFP$_2$. We want to augment this with results up to GFP$_5$ and compare the resulting simulations with the transport solution. We focus on the following problem:
\begin{align}
\sigma_a \Psi(z,\mu) + \mu \D{\Psi(z,\mu)}{z} = \mu_s \left[ \int_{4\pi} \sigma_s(\ul{\Omega}\cdot\ul{\Omega}')\Psi(z,\ul{\Omega}') d\Omega' -\Psi(z,\mu) \right] \label{DE_RodKim}
\end{align}
\begin{align*}
	\ul{\text{BC}}:& \quad \Psi(z=0,\mu) =e^{-10(1-\mu)^2} \hspace{3.7cm}  1\geq \mu>0 \\
								 & \quad \Psi(z=d=2,\mu) =0  \hspace{4cm} -1\leq \mu<0
\end{align*}
It is a slab geometry with a thickness of $d=$ 2mm disregarding any time dependence. Its solution enables to compute reflectance $R(\mu)$ and transmittance $T(\mu)$ defined by
\begin{align*}
	R(\mu)&=\Psi(\mu,0) \qquad -1 \leq \mu < 0 \\
	T(\mu)&=\Psi(\mu,d) \hspace{1.1cm} 1 \geq \mu > 0. \\
\end{align*}
\begin{figure}%
\centering \includegraphics[scale=0.6]{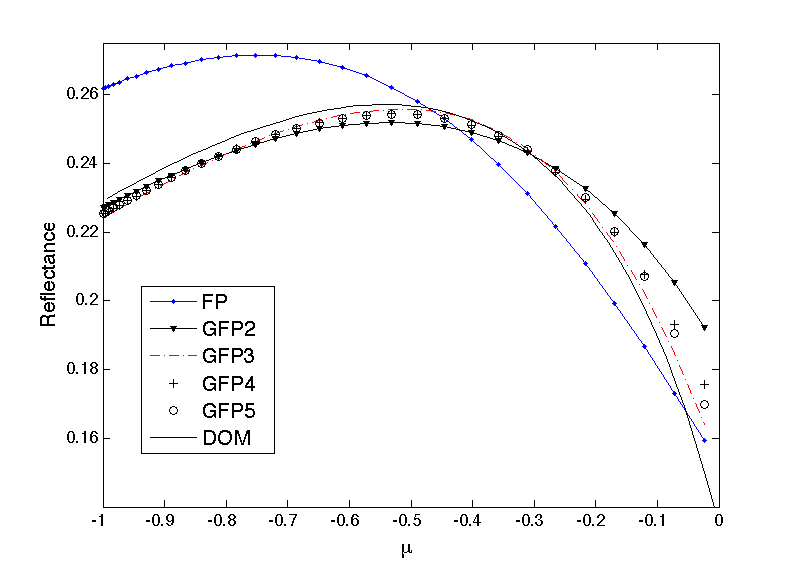} \\
\centering \includegraphics[scale=0.65]{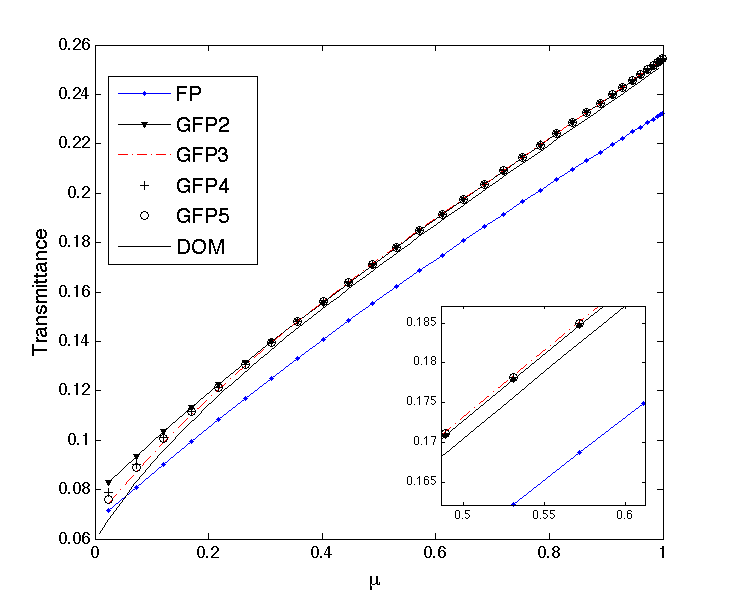}%
\setcaptionwidth{.88\textwidth}
\caption{Single HG with g=0.98: Reflectance and transmittance of liver tissue arising from a slab geometry with thickness d=2mm and conditions in \refeq{DE_RodKim}. Transmittance is plotted in a semilogarithmic scale.} \label{HG_RodKim}%
\end{figure}%
\begin{figure}%
\centering \includegraphics[scale=0.6]{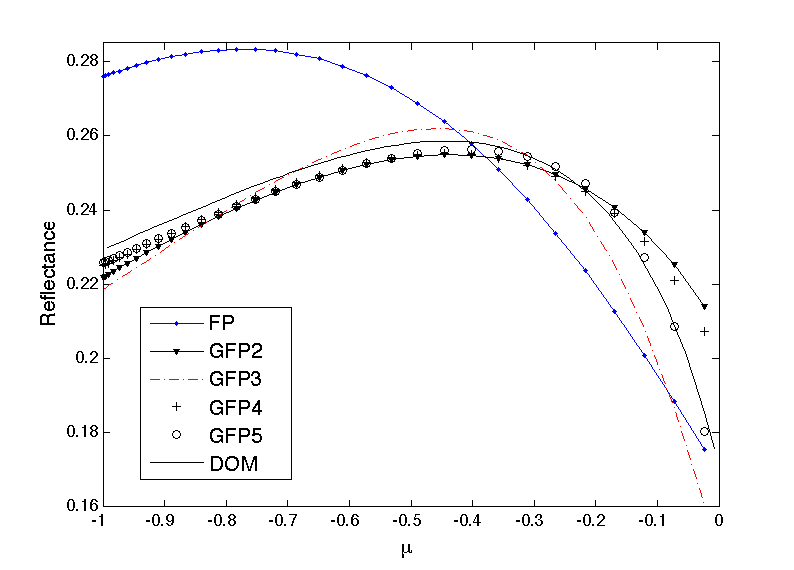} \\
\centering \includegraphics[scale=0.6]{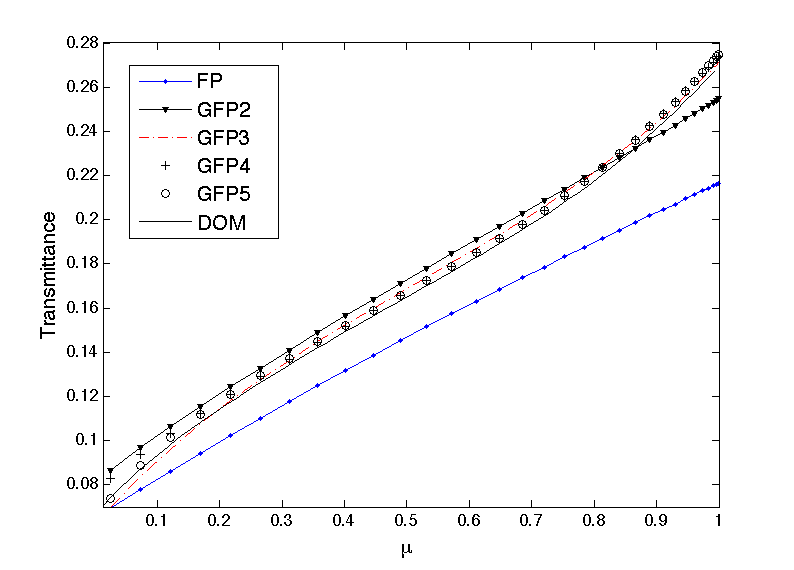}%
\setcaptionwidth{.86\textwidth}
\caption{Double HG with $g_1=0.85$, $g_2=-0.34$: GFP approximations for reflectance and transmittance of liver tissue. Transmittance is plotted in a semilogarithmic scale.} \label{DHG_RodKim}%
\end{figure}%
\noindent A. \quad SINGLE HENYEY-GREENSTEIN KERNEL \\ \\
In the first run \refeq{DE_RodKim} was solved with discretizations of 64 points in angle and 80 points in space using GFP approximations with the HG DSCS $\sigma_s=\sigma_s^{HG}$. Further constants were set to
\[ g=0.98 \quad \sigma_a=0.01 \text{mm}^{-1} \quad \mu_s=50 \text{mm}^{-1}. \]

\noindent REFLECTANCE: Starting at $R(-1)\approx 0.23$ the reflectance slightly increases and attains its maximum at $ \mu \approx -0.5$ (\reffig{HG_RodKim}). Although in this interval GFP data show discrepancies among each other their results are accurate and GFP$_3$ gives the best approximation. For $\mu>-0.5$ the transport solution DOM hunches down more than GFP functions and hence, the error increases rapidly. Surprisingly, for $\mu \gtrsim -0.25$ GFP$_3$-reflectance values are closer to DOM than those of GFP$_5$. Throughout the whole interval FP values give a very poor approximation.

TRANSMITTANCE: It is almost a straight line only bending for small $\mu$. In contrast to the reflectance, a more or less constant distance to the transport solution is always sustained. To the eye, there are no differences between all GFP simulations in a wide range. Only for small $\mu$ the functions start to deviate and GFP$_3$ data give best results whereas FP is inaccurate again.\\ \\
\noindent B. \quad DOUBLE HENYEY-GREENSTEIN KERNEL \\ \\
Taking the amount of large-angle scattering in biological tissue into account Gonz\'a lez-Rodr\'iguez and Kim applied the double Henyey-Greenstein DSCS to simulate transmittance and reflectance in liver tissue. The following fit parameters were used:
\[ g_1=0.85 \quad g_2=-0.34 \quad b=0.86. \]
$g_1=0.85$ provides a forward-peak which is not very sharp. In addition, the combination of $g_2=-0.34$ and $b=0.86$ contains a significant amount of large-angle scattering which leads to increasing STCs:
\[ \xi_1=0.3166 \quad  \xi_2=0.3916  \quad  \xi_3=0.6058  \quad  \xi_4=1.0075  \quad  \xi_5=1.7388 \]
In this case our fundamental assumption is not valid which could negatively affect our approximations. Moreover, it is important to emphasize that simulations for GFP$_3$-GFP$_5$ ran with some \emph{negative} coefficients $\alpha_i$, $\beta_i$.  Nevertheless, our code gave reasonable results plotted in \reffig{DHG_RodKim} for discretization parameters of 64 points in $\mu$ and 70 points in $z$ direction.

REFLECTANCE: Fig.~\ref{DHG_RodKim} shows 'bump head' functions similarly shaped to those of the single HG kernel. The x-coordinates of their maxima are, however, shifted to the right. Moreover, for large $\mu$ different GFP approximations do not match as well as they do in \reffig{HG_RodKim}. In contrast to the single HG kernel GFP$_3$ data give a poor approximation whereas GFP$_5$ is the best one among all shown here. Only for $\mu \approx 0$ GFP$_2$ is not able to match GFP$_5$. As expected, FP gives even worse results than for the single HG kernel. 

TRANSMITTANCE: This time our transport solution DOM is more peaked at $\mu \approx 1$. Nevertheless GFP gives more accurate results than in \reffig{HG_RodKim}. A comparison between the two best approximations GFP$_3$ and GFP$_5$ yields small differences which enlarge near $\mu \approx 0$. However, the classical FP deviates from our benchmark to a big extent.

\noindent Due to the contribution of large angle scattering numerical computations with the double HG kernel are more challenging and, in fact, give GFP coefficients which contradict our assumptions. Nevertheless, the GFP results plotted above approximate the transport solution very well and are much more precise than those of the FP calculations.
%
%
%
%
\subsection{Slab Geometry: Electron Propagation in Tissue}
For dose calculations the following GFP equation is to be solved (examplarily stated for GFP$_2$):
\begin{align}
\sigma_a \Psi^{(0)}(z,E,\mu) + \D{\Psi^{(0)}(z,E,\mu)}{z} \cdot \mu &= \alpha L_{\mu} \Psi^{(1)}(z,E,\mu) + \frac{\partial (S(z,E) \Psi^{(0)}(z,E,\mu))}{\partial E} \nonumber \\
(I-\beta L_{\mu})\Psi^{(1)}(z,\mu,s) &= \Psi^{(0)}(z,E,\mu) \label{PomGFP2}
\end{align}
\begin{align*}
	\ul{\text{BC}}:& \quad \Psi^{(0)}(0,E,\mu) =10^5 \cdot e^{-200(1-\mu)^2} e^{-50(E_0-E)^2} \hspace{1.0cm}  1\geq \mu>0, E \in I .\\ \nonumber
								 & \quad \Psi^{(0)}(d,E,\mu) =0  \hspace{4.7cm} -1\leq \mu<0, E \in I.
\end{align*}
The initial boundary value problem in \refeq{PomGFP2} describes the propagation of electrons through matter with a monoenergetic pencil beam of energy $E_0$ irradiated orthogonally to the boundary surface of the material. This is modelled by a product of two narrow Gaussian functions around $\mu=1$ and $E=E_0$. After computing the solution one can calculate the absorbed dose via
\begin{align}
D(\ul{r}) &= \frac{2\pi T}{\rho(\ul{r})} \int_0^\infty \int_{-1}^{1} S(\ul{r},E') \Psi^{(0)}(\ul{r},\mu,E') d\mu dE'.
\end{align}
$T$ is hereby the duration of the irradiation of the patient and $\rho$ the mass density of the irradiated tissue so that $D(\ul{r})$ leads to SI unit $J/kg$ or $Gy$.

\noindent Several test cases were implemented for 5 MeV and 10 MeV beams. As benchmark we used solutions of the MC code systems GEANT4 (standard physics package) \cite{AgoAllAma03, AllAmaApo06} and PENELOPE \cite{Pen09}. However, it should be stressed that all physical models were obtained independently. The following criteria are generally employed to quantify the accuracy of a dose curve \cite{VenWelMij01}: 2\%/2mm (pointwise difference within 2\% or 2mm horizontal distance-to-agreement) in homogeneous and 3\%/3mm in inhomogeneous geometries. \\

\begin{figure}%
\centering \includegraphics[scale=0.5]{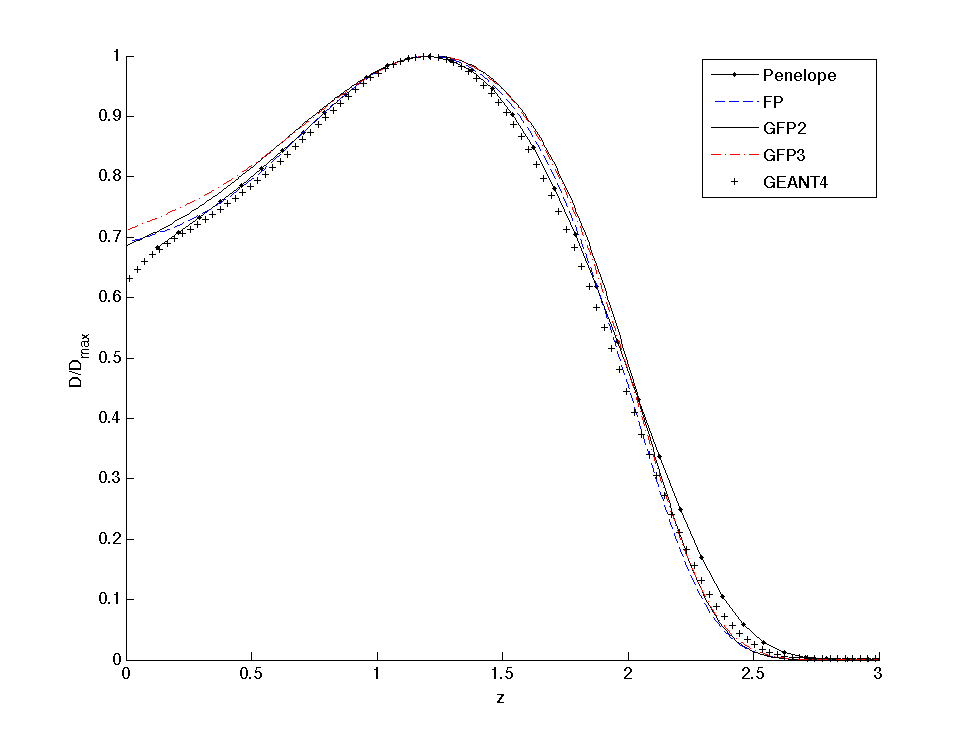}%
\caption{Normalized dose in liquid water for a 5 MeV electron beam.} \label{5MeVelectron}%
\centering \includegraphics[scale=0.5]{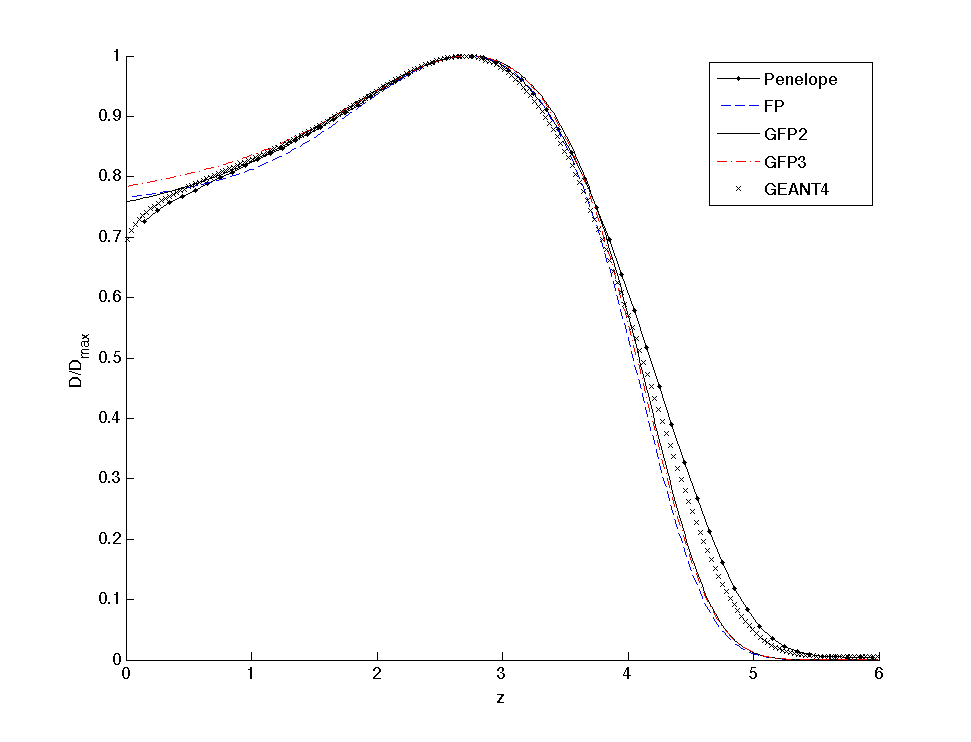}%
\caption{10 MeV electron beam: normalized dose in liquid water.} \label{10MeVelectron}%
\end{figure}

\noindent A. \quad HOMOGENEOUS GEOMETRY \\ \\
Characteristic electron dose profiles in a semi-infinite water phantom are shown in \reffig{5MeVelectron} and \reffig{10MeVelectron}. First they provide a high surface dose, increase to a maximum at a certain depth and drop off with a steep slope afterwards. Solutions for GFP$_4$ and GFP$_5$ are omitted because they overlap with GFP$_3$ in our plot. Except for GFP$_2$, computations were performed according to \refeq{morel} (32 points in $\mu$, 350 points in $z$). Due to better results we applied upwind finite difference discretizations for GFP$_2$, equidistant in $z$ (400 points) and $\mu$ (200 points). All approximations are close to each other because GFP transport coefficients $\xi_n(E)$ for water do not fall off highly enough within our energy interval. All in all, the calculated results agree well with PENELOPE and GEANT4. All dose profiles for a 5 MeV beam satisfy the 2\%/2mm criterion. As we neglect bremsstrahlung the difference to MC computations becomes bigger for $10$ MeV. In fact, the largest FP and GFP$_2$ distance to PENELOPE and GEANT4 becomes 3mm at $z\approx 5$ cm and hence, they do not meet the criterion. \\

\noindent B. \quad INHOMOGENEOUS GEOMETRIES \\ \\
Dose calculation is more challenging in parts of the body where materials of strongly varying densities meet. Here, large dosimetric differences between experiments and predictions exist \cite{MarReyWag02}. As deviations of already five percent in the deposited dose may result in a 20\% to 30\% impact on complication rates \cite{AAPMreport85} it is of big importance to accurately compute the dose in such transition regions.

\begin{figure}%
\centering \includegraphics[scale=0.65]{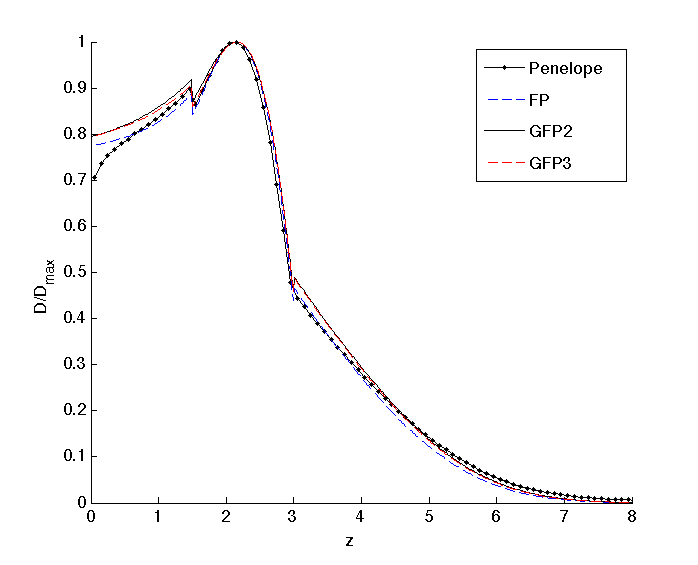}%
\caption{Normalized dose curves of 10 MeV electrons irradiated on the back of the body (32 points in $\mu$, 300 points in $z$).} \label{10MeVback}%
\centering \includegraphics[scale=0.65]{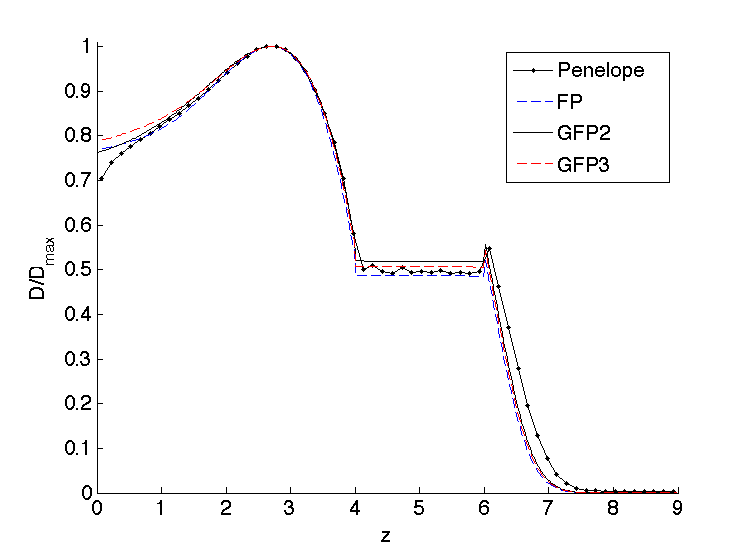}%
\caption{10 MeV electron beam: normalized dose in liquid water with air cavity.} \label{10MeVcavity}%
\end{figure}

Possible clinical applications for electron beams are for example irradiation of the chest wall or the vertebral column. To simulate dose curves on the central beam we assume that 10 MeV electrons pass three different materials: muscle (0-1.5cm), bone (1.5-3cm) and lung (3-9cm). For all results, parameters for Morel's discretization \cite{Mor85} were set to 32 points in $\mu$ and 400 points in $z$. Fig.~\ref{10MeVback} illustrates approximations up to order three because higher order results overlap with the latter on that scale. The agreement with the MC dose profile is very satisfactory although bigger differences occur for small penetration depths. The dose differences between PENELOPE and FP exceed the 3\%/3mm limit only at the boundary $z=0$.

Radiotherapy gains in importance not least because surgical interventions can be avoided. Especially sensitive body areas like the brain, coated by cerebral membranes, are of big interest. Between those membranes there are many voids which means that scattering and absorption properties change abruptly. Therefore we consider an air cavity irradiated by a 10 MeV electron beam first penetrating water (0-4cm), then air (4-6cm) and later water (6-9cm) again. Similar to pure water, GFP$_2$ calculations yield better solutions for equidistant upwind discretizations (200 points in $\mu$ and 300 points in $z$). Remaining curves were obtained by Morel's scheme ($\mu$-direction: 32 points, $z$-direction: 350 points).  Except for small penetration depths all deterministic solutions are very close to each other and demonstrate good aggreement with PENELOPE (\reffig{10MeVcavity}). Again FP and GFP$_2$ results show the best approximations. Larger, but still comparably small, differences between them occur in the air region. Except for the boundary value ($z=0$), the FP and GFP$_2$ curves fulfill the 3\%/3mm criterion. \\

%
%
%
%
\section{Conclusions}
\noindent We studied practical applications of Generalized Fokker Planck approximations. Numerical examples of GFP solutions for the Henyey-Greenstein kernel in slab geometry showed more accurate approximations than FP calculations. Further test cases for reflectance and transmittance in liver tissue by means of single and doulbe HG kernels also revealed GFP$_3$- and GFP$_5$-results closest to the transport solution.

\noindent For electron transport, we derived an \textit{ab initio} model from the ICRU database. This model was compared to publicly available MC Codes (PENELOPE and GEANT4) which in turn has been benchmarked against experiments. We extracted the stopping power, elastic and inelastic cross sections from the ICRU database and transformed them to transport coefficients needed for GFP computations. Dose distributions for electron beams were performed without additional coupling to photons and positrons. We are aware that our physical model neglects important interactions like energy straggling and hard radiative events with emission of photons. They are inevitable for accurate dose calculations with high-energy electrons. However, in our energy range they are less frequent and regarded as extensions for improved models in future. And in fact, comparisons of GFP approximations with Monte Carlo calculations reveal dose profiles which agree well in both homogeneous and inhomogeneous geometries. 

\noindent Several tasks for further examination remain:
\begin{enumerate}[(i)]

\item The first step towards real dose calculations from CT data is an extension to two space dimensions. As the GFP theory was derived for angular fluxes in 3D space this should only be a challenge to the numerical and programming approach.

\item Due to a rising demand for proton therapy facilities the adaption of the GFP theory to protons is certainly an interesting subject of further study.

\item To improve computational results it is necessary to include more physical phenomena. Especially for high energy electron beams it is inevitable to simulate the transport of bremsstrahlung quanta.  
\end{enumerate}
%
%
%
%
\section*{Acknowledgements}
\noindent The authors would like to thank Bruno Dubroca from Universit\'e Bordeaux not only for providing his code for the discrete ordinates method but also for his support and advice. We also acknowledge support from the German Research Foundation DFG under grant KL 1105/14/2 and the German Academic Exchange Sevice DAAD Program D/07/07534.
%
%
%
%
%
%
%
\begin{appendix}
\section{Polynomial Operators}
\begin{align}
L_{P2} &= \left (\frac{\xi_1}{2} +\frac{\xi_2}{8} \right )L + \left (\frac{\xi_2}{16} \right )L^2 \\
L_{P3} &= \left (\frac{\xi_1}{2} +\frac{\xi_2}{8} +\frac{\xi_3}{24} \right )L + \left (\frac{\xi_2}{16} +\frac{\xi_3}{36} \right )L^2 + \left (\frac{\xi_3}{288} \right )L^3 \label{LP3} \\
L_{P4} &= \left (\frac{\xi_1}{2} +\frac{\xi_2}{8} +\frac{\xi_3}{24} +\frac{\xi_4}{64} \right )L + \left (\frac{\xi_2}{16} +\frac{\xi_3}{36} +\frac{3\xi_4}{256} \right )L^2 \nonumber \\
	& \quad + \left (\frac{\xi_3}{288} +\frac{5\xi_4}{2304} \right )L^3 + \left (\frac{\xi_4}{9216} \right )L^4 \label{LP4}
\end{align}
\begin{align}
L_{P5} &= \left (\frac{\xi_1}{2} +\frac{\xi_2}{8} +\frac{\xi_3}{24} +\frac{\xi_4}{64} +\frac{\xi_5}{160} \right )L + \left (\frac{\xi_2}{16} +\frac{\xi_3}{36} +\frac{3\xi_4}{256} +\frac{\xi_5}{200} \right )L^2 \nonumber \\
	& \quad + \left (\frac{\xi_3}{288} +\frac{5\xi_4}{2304} +\frac{127\xi_5}{115200} \right )L^3 + \left (\frac{\xi_4}{9216} +\frac{\xi_5}{11520} \right )L^4 \nonumber \\
	& \quad +\frac{\xi_5}{460800}L^5 \label{LP5}
\end{align}
%
%
%
%
%
%
\section{Derivation of Generalized Fokker-Planck Operators}
\begin{center} \eb{GFP$_2$} \end{center}
If $\lambda_n^{B}$ denotes one eigenvalue of $L_{B}$ for $n\geq 0$ and $\alpha$, $\beta$ are two \textit{positive} constants the Generalized Fokker-Planck (GFP) operator defined by
\begin{align}
	L_{P2} + \mathcal{O}(\eps^2) =	L_{GFP_2} & :=\alpha L(I-\beta L)^{-1} \nonumber \\
 & = \alpha L + \alpha\beta L^2 + {\cal O}(\alpha\beta^2)
\end{align}
will have to satisfy three properties to substitute $L_{P2}$ in the favoured way:
\begin{enumerate}
	\item{Eigenvalue preservation
				\begin{align*}
					-\frac{\alpha n(n+1)}{1+\beta n(n+1)} & = \lambda^{GFP_2}_n \nach{!} \lambda^B_{n} = -\sigma_{an} \quad \text{for} \quad n=1,2 \nonumber
				\intertext{Multiplying above equation by $(1+\beta n(n+1)) \neq 0$ and dividing by n(n+1) we conclude:}
				(\alpha - \beta\sigma_{an}) &= \frac{\sigma_{an}}{n(n+1)} \quad n=1,2 \nonumber \\
				\Leftrightarrow \begin{bmatrix}
												1 & -\sigma_{a1} \\
												1 & -\sigma_{a2}
												\end{bmatrix}
												\cdot
												\begin{bmatrix}
												\alpha \\
												\beta
												\end{bmatrix}
												& = \begin{bmatrix}
													\sigma_{a1}/2 \\
													\sigma_{a2}/6
													\end{bmatrix}
				\end{align*}
			 }
	\item{Order
					$\q \q {\cal O}(\alpha\beta^2) \nach{!} {\cal O}(\eps^2)$
			 }
\item{Equivalence
				\begin{align}
					\alpha L +\alpha\beta L^2 & \nach{!} L_{P2} + {\cal O}(\eps^2) \nonumber \\
					& = \left( \frac{\xi_1}{2} + \frac{\xi_2}{8} \right) L + \frac{\xi_2}{16} L^2 + {\cal O}(\eps^2). \nonumber
				\end{align}
			 }	 
\end{enumerate}
$\sigma_{an}$ is a quantity which can be expressed in terms of $\xi_n$:
\begin{align}
\sigma_{a0}  &= 0    \label{Sigma_s=xi_0} \\
\sigma_{a1} &= \xi_1    \label{sigma_a1} \\
\sigma_{a2} &= 3\xi_1 -\frac{3}{2}\xi_2        \label{sigma_a2} \\
\sigma_{a3} &= 6\xi_1 -\frac{15}{2}\xi_2 +\frac{5}{2}\xi_3           \label{sigma_a3} \\
\sigma_{a4} &= 10\xi_1 -\frac{45}{2}\xi_2 +\frac{35}{2}\xi_3 -\frac{35}{8}\xi_4             \label{sigma_a4} \\
\sigma_{a5} &= 15\xi_1 -\frac{105}{2}\xi_2 +70\xi_3 -\frac{315}{8}\xi_4 +\frac{63}{8}\xi_5.        \label{sigma_a5}
\end{align}
In the following item (1) is first transformed to a system of linear equations and thereafter solved for the desired GFP coefficients (here: $\alpha$ and $\beta$). However, the final equations to be solved are non-linear for GFP operators of order $n\geq3$. In this case of order $n=2$ eqs.~(\ref{sigma_a1})-(\ref{sigma_a2}) yield
\begin{align*}
	\alpha = \frac{\xi_1}{2} + \frac{\xi_2}{8} \q \tx{and} \q \beta = \frac{\xi_2}{8\xi_1}.
\end{align*}
Going on with item (2) it is to be checked:
\begin{align*}
	\alpha \beta^2 = \left( \frac{\xi_1}{2} + \frac{\xi_2}{8} \right) \left( \frac{\xi_2}{8\xi_1} \right)^2 = \underbrace{\left( \frac{\xi_1}{2} + \frac{\xi_2}{8} \right)}_{\makebox[0pt]{$\begin{array}{c} \in {\cal O}(1) \end{array}$}} \underbrace{\left( \frac{\xi_2^2}{64\xi_1^2} \right)}_{\makebox[0pt]{$\begin{array}{c} \in {\cal O}(\eps^2)  \end{array}$}} \in {\cal O}(\eps^2) \quad \surd 
\end{align*}
As equivalence condition follows straight forward it has been shown that the operator $L_{GFP_2}$ is an ${\cal O}(\eps^2)$ approximation to $L_B$ whose first three eigenvalues agree.

\noindent Now it is quite intuitive to apply a similar procedure to higher order operators. We determined explicit solutions for GFP$_2$-GFP$_5$ coefficients and performed verifications for items (1)-(3). According to GFP$_3$ all items were checked without computer support. The asymptotic behaviour of GFP operators of order four and five was, however, checked by means of a symbolic toolbox. To keep our following description short we confine ourselves to final results.
\begin{center} \eb{GFP$_3$} \end{center}
\begin{align}
  L_{P3} + \mathcal{O}(\eps^3) = L_{GFP_3} :=& \alpha_1 L(I-\beta_1 L)^{-1} +\alpha_2 L \nonumber \\
	=& (\alpha_1 +\alpha_2)L +\alpha_1\beta_1 L^2 +\alpha_1\beta_1^2 L^3 +{\cal O}(\alpha_1\beta_1^3) \label{GFP3}
\end{align}
\begin{enumerate}
	\item{Eigenvalue preservation
				\begin{align}
				& (\alpha_1 +\alpha_2) -\sigma_{an}\beta_1 +n(n+1)\beta_1\alpha_2 = \frac{\sigma_{an}}{n(n+1)} & \quad n=1,2,3 \label{linsys3} \\
				\Rightarrow \,
				\alpha_1 &= \frac{\xi_2(27\xi_2^2 +5\xi_3^2 -24\xi_2\xi_3)}{8\xi_3(3\xi_2 -2\xi_3)} & \label{alpha1} \\
				\beta_1  &= \frac{\xi_3}{6(3\xi_2 -2\xi_3)}  &  \label{beta1} \\
				\alpha_2 &= \frac{\xi_1}{2} -\frac{9\xi_2^2}{8\xi_3} +\frac{3\xi_2}{8} &  \label{alpha2}
				\end{align}
				}
 \item[(3)]{Equivalence \\
 			\smallskip
				To guarantee that $L_{GFP_3}=L_{P3} +{\cal O}(\eps^3)$ all coefficients of $L^i$ in  \refeq{GFP3} and \refeq{LP3} must coincide. Let $\alpha_1, \beta_1, \alpha_2$ be the defining \textit{positive} coefficients of $L_{GFP_3}$ as stated in eqs.  (\ref{alpha1})-(\ref{alpha2}) then one can show that they satisfy
				\begin{enumerate}[I.] 
				 \item{$\alpha_1 +\alpha_2 = \ds \frac{\xi_1}{2} +\frac{\xi_2}{8} +\frac{\xi_3}{24} +{\cal O}(\eps^3)$}		
				 \item{$\alpha_1\beta_1 = \ds \frac{\xi_2}{16} +\frac{\xi_3}{36} +{\cal O}(\eps^3)$}
				 \item{$\alpha_1\beta_1^2 = \ds \frac{\xi_3}{288} +{\cal O}(\eps^3)$}.
				\end{enumerate}
				}
\end{enumerate}
\medskip
\begin{center} \eb{GFP$_4$} \end{center}
\begin{align}
		L_{P4} + \mathcal{O}(\eps^4) = L_{GFP_4} :=& \alpha_1 L(I-\beta_1 L)^{-1} +\alpha_2 L(I-\beta_2L)^{-1} \nonumber \\
	=& (\alpha_1 +\alpha_2)L + (\alpha_1\beta_1 +\alpha_2\beta_2) L^2 +(\alpha_1\beta_1^2 +\alpha_2\beta_2^2) L^3 \nonumber \\ 
	+& (\alpha_1\beta_1^3 +\alpha_2\beta_2^3) L^4  +{\cal O}(\alpha_1\beta_1^4) +{\cal O}(\alpha_2\beta_2^4) \label{GFP4}
\end{align}
\begin{enumerate}
	\item{Eigenvalue preservation
				\begin{align}
				(\alpha_1 +\alpha_2) -& \sigma_{an}(\beta_1 +\beta_2) -\sigma_{an}n(n+1)\beta_1\beta_2 \nonumber \\
				+& n(n+1)(\alpha_1\beta_2 +\alpha_2\beta_1) = \frac{\sigma_{an}}{n(n+1)} \quad n=1,2,3,4 \label{linsys4}
				\end{align}
				}
		\item[(3)]{Equivalence
 			\begin{enumerate}[I.]
 				\item{ $\alpha_1 +\alpha_2 = \ds \frac{\xi_1}{2} +\frac{\xi_2}{8} +\frac{\xi_3}{24} +\frac{\xi_4}{64} +{\cal O}(\eps^4)$
 						 }
 				\item{ $\alpha_1\beta_1 +\alpha_2\beta_2 = \ds \frac{\xi_2}{16} +\frac{\xi_3}{36} +\frac{3\xi_4}{256} +{\cal O}(\eps^4)$
 						 }
 				\item{ $\alpha_1\beta_1^2 +\alpha_2\beta_2^2 = \ds \frac{\xi_3}{288} +\frac{5\xi_4}{2304} +{\cal O}(\eps^4)$
 						 }
 				\item{ $\alpha_1\beta_1^3 +\alpha_2\beta_2^3 = \ds \frac{\xi_4}{9216} +{\cal O}(\eps^4)$
 						 }
 			\end{enumerate}
			   }
\end{enumerate}
\medskip
\begin{center} \eb{GFP$_5$} \end{center} 
\begin{align}
	L_{P5} + \mathcal{O}(\eps^5) = L_{GFP_5} :=& \alpha_1 L(I-\beta_1 L)^{-1} +\alpha_2 L(I-\beta_2L)^{-1} +\alpha_3L \nonumber \\
=& (\alpha_1 +\alpha_2 +\alpha_3)L +(\alpha_1\beta_1 +\alpha_2\beta_2) L^2 +(\alpha_1\beta_1^2 +\alpha_2\beta_2^2) L^3  \nonumber \\ 
	 +& (\alpha_1\beta_1^3 +\alpha_2\beta_2^3) L^4 +(\alpha_1\beta_1^4 +\alpha_2\beta_2^4) L^5  +{\cal O}(\alpha_1\beta_1^5) +{\cal O}(\alpha_2\beta_2^5) \label{GFP5}
\end{align}
\begin{enumerate}
	\item{Eigenvalue preservation
				\begin{align}
				(\alpha_1 +\alpha_2 +\alpha_3) -& \sigma_{an}(\beta_1 +\beta_2) -\sigma_{an}n(n+1)\beta_1\beta_2  \nonumber \\
				 +& n(n+1)\left[ \beta_2(\alpha_1 +\alpha_3) +\beta_1(\alpha_2 +\alpha_3) \right] \nonumber \\
				 +& \alpha_3\beta_1\beta_2 \left[ n(n+1) \right]^2 = \frac{\sigma_{an}}{n(n+1)} \quad n=1,2,3,4,5 \label{linsys5}
				\end{align}
				}
		\item[(3)]{Equivalence
 			\begin{enumerate}[I.]
 				\item{ $\alpha_1 +\alpha_2 +\alpha_3 = \ds \frac{\xi_1}{2} +\frac{\xi_2}{8} +\frac{\xi_3}{24} +\frac{\xi_4}{64} +\frac{\xi_5}{160} +{\cal O}(\eps^5)$
 						 }
 				\item{ $\alpha_1\beta_1 +\alpha_2\beta_2 = \ds \frac{\xi_2}{16} +\frac{\xi_3}{36} +\frac{3\xi_4}{256} +\frac{\xi_5}{200} +{\cal O}(\eps^5)$
 						 }
 				\item{ $\alpha_1\beta_1^2 +\alpha_2\beta_2^2 = \ds \frac{\xi_3}{288} +\frac{5\xi_4}{2304} +\frac{127\xi_5}{115200} +{\cal O}(\eps^5)$
 						 }
 				\item{ $\alpha_1\beta_1^3 +\alpha_2\beta_2^3 = \ds \frac{\xi_4}{9216} +\frac{\xi_5}{11520}  +{\cal O}(\eps^5)$
 						 }
 				\item{ $\alpha_1\beta_1^4 +\alpha_2\beta_2^4 = \ds \frac{\xi_5}{460800} +{\cal O}(\eps^5)$
 						 }
 			\end{enumerate}
			   }
\end{enumerate} 
\medskip
All linear and nonlinear equations stated above lead to explicit solutions for $\alpha_i$ and $\beta_i$. Equations posed for GFP$_2$ and GFP$_3$ actually deliver \textit{uniquely} determined constants whereas for higher GFP operators there is no guarantee for unique or even real valued $\alpha_i$ and $\beta_i$. Nevertheless, it is important to emphasize that the resulting values of $\alpha_i$ and $\beta_i$ must be positive. Otherwise, eigenvalues $\lambda_{n}^{GFP_k}$ of a GFP$_k$ operator could become negative. For DSCS $\sigma_s(\ul\Omega\cdot\ul\Omega')$ of different materials, and thus different $\xi_n$, this has to be checked separately.
\end{appendix}
%
%
%
%
%
%

\nocite{*}

\bibliography{literature}

\begin{thebibliography}{10}

\bibitem{AgoAllAma03}
{\sc S.~Agostinelli et~al.}, {\em Geant4--a simulation toolkit}, Nucl. Instrum.
  Methods B, 506 (2003), pp.~250--303.

\bibitem{AhnSaxTre92}
{\sc A.~Ahnesj\"o, M.~Saxner, and A.~Trepp}, {\em A pencil beam model for
  photon dose calculation}, Med. Phys., 19 (1992).

\bibitem{AllAmaApo06}
{\sc J.~Allison et~al.}, {\em Geant4 developments and applications}, IEEE
  Transactions on Nuclear Science, 53 (2006), pp.~270--278.

\bibitem{And91}
{\sc P.~Andreo}, {\em Monte {C}arlo techniques in medical radiation physics},
  Phys. Med. Biol., 36 (1991), pp.~861--920.

\bibitem{AydOliGod02}
{\sc E.~D. Aydin, C.~R.~E. de~Oliveira, and A.~J.~H. Goddard}, {\em A
  comparison between transport and diffusion calculations using finite
  element-spherical harmonics radiation transport method}, Med. Phys., Vol. 29,
  No. 9 (2002), pp.~2013--2023.

\bibitem{Bal00}
{\sc D.~Balsara}, {\em Fast and accurate discrete ordinates methods for
  multidimensional radiative transfer. {P}art {I}, basic methods}, Journal of
  Quantitative Spectroscopy \& Radiative Transfer, 69 (2001), pp.~671--707.

\bibitem{BomTerVau05}
{\sc E.~Boman, J.~Tervo, and M.~Vauhkonen}, {\em Modelling the transport of
  ionizing radiation using the finite element method}, Phys. Med. Biol., 50
  (2005), pp.~265--280.

\bibitem{Bor98}
{\sc C.~B\"orgers}, {\em Complexity of {M}onte {C}arlo and deterministic dose
  calculation methods}, Phys. Med. Biol., 43 (1998), pp.~517--528.

\bibitem{Bru02}
{\sc T.~A. Brunner}, {\em Forms of application radiation transport}, {S}andia
  {R}eport, sand2002-1778, Sandia National Laboratories, 2002.

\bibitem{ChePraWel90}
{\sc W.~Cheong, S.~A. Prahl, and A.~J. Welch}, {\em A review of the optical
  properties of biological tissues}, IEEE Journal of Quantum Electronics, 26
  (1990), pp.~2166--2185.

\bibitem{CopRav95}
{\sc G.~G.~M. Coppa and P.~Ravetto}, {\em Quasi-singular angular finite element
  methods in neutron transport problems}, Trans. Theory Stat. Phys., 24 (1995),
  pp.~155--172.

\bibitem{CygLocDas05}
{\sc J.~Cygler et~al.}, {\em Clinical use of a commercial monte carlo treatment
  planning system for electron beams}, Phys. Med. Biol., 50 (2005).

\bibitem{DatAltRay96}
{\sc R.~P. Datta et~al.}, {\em Computational model for coupled electron-photon
  transport in two dimensions}, Phys. Rev., E 53 (1996), pp.~6514--6522.

\bibitem{DucMorTik09}
{\sc R.~Duclous et~al.}, {\em Reduced multi-scale kinetic model for the
  relativistic electron transport in solid targets: effects related to
  secondary electrons}, Laser and Particle Beams,  (2010).

\bibitem{Eds05}
{\sc P.~Edstr\"om}, {\em A fast and stable solution method for the radiative
  transfer problem}, SIAM Review, 47 (2005), pp.~447--468.

\bibitem{Eyg48}
{\sc L.~Eyges}, {\em Multiple scattering with energy loss}, Phys. Rev., 74
  (1948), pp.~1534--1535.

\bibitem{Fer40}
{\sc E.~Fermi}, {\em The ionization loss of energy in gases and in condensed
  materials}, Phys. Rev., 57 (1940), pp.~485--493.

\bibitem{FraHenKla07}
{\sc M.~Frank, H.~Hensel, and A.~Klar}, {\em A fast and accurate moment method
  for the {F}okker-{P}lanck equation and applications to electron
  radiotherapy}, SIAM J. Appl. Math., 67 (2007), pp.~582--603.

\bibitem{FraHerSan09}
{\sc M.~Frank, M.~Herty, and A.~N. Sandjo}, {\em Optimal treatment planning
  governed by kinetic equations}, to appear in Math. Mod. Meth. Appl. Sci.,
  (2009).

\bibitem{FraHerSch08}
{\sc M.~Frank, M.~Herty, and M.~Sch{\"a}fer}, {\em Optimal treatment planning
  in radiotherapy based on {B}oltzmann transport calculations}, Math. Mod.
  Meth. Appl. Sci., 18 (2008), pp.~573--592.

\bibitem{Gar04}
{\sc J.~C. Garth}, {\em Electron/photon transport - a key technology to
  radiation physics? a review, invited}, Transactions of the American Nuclear
  Society,  (2004), pp.~289--291.

\bibitem{Gar05}
\leavevmode\vrule height 2pt depth -1.6pt width 23pt, {\em Electron/photon
  transport and its applications}, The Monte Carlo Method: Versatility
  Unbounded in a Dynamic Computing World,  (2005).

\bibitem{GifHorWar06}
{\sc K.~A. Gifford et~al.}, {\em Comparison of a finite-element multigroup
  discrete-ordinates code with {M}onte {C}arlo for radiotherapy calculations},
  Phys. Med. Biol., 51 (2006), pp.~2253--2265.

\bibitem{RodKim08}
{\sc P.~Gonz\'alez-Rodr\'iguez and A.~D. Kim}, {\em Light propagation in
  tissues with forward-peaked and large-angle scattering}, Appl. Opt., Vol. 47,
  No.14 (2008), pp.~2599--2609.

\bibitem{HenIzaSie06}
{\sc H.~Hensel, R.~Iza-Teran, and N.~Siedow}, {\em Deterministic model for dose
  calculation in photon radiotherapy}, Phys. Med. Biol., Vol. 51 (2006),
  pp.~675--693.

\bibitem{HenGre41}
{\sc L.~G. Henyey and J.~L. Greenstein}, {\em Diffuse radiation in the galaxy},
  Astrophysics Journal, vol. 93 (1941), pp.~70--83.

\bibitem{HogMilAlm81}
{\sc K.~R. Hogstrom, M.~D. Mills, and P.~R. Almond}, {\em Electron beam dose
  calculations}, Phys. Med. Biol., 26 (1981), pp.~445--459.

\bibitem{HuiSto89}
{\sc H.~Huizenga and P.~Storchi}, {\em Numerical calculation of energy
  deposition by broad high-energy electron beams}, Phys. Med. Biol., 34 (1989),
  pp.~1371--1396.

\bibitem{ICRU07}
{\sc ICRU}, {\em Elastic scattering of electrons and positrons, report 77},
  Journal of the ICRU, 7 (2007).

\bibitem{ItiMas05}
{\sc Y.~Itikawa and N.~Mason}, {\em Cross sections for electron collisions with
  water molecules}, J. Phys. Chem. Ref. Data, 34 (2005).

\bibitem{JanRieMor94}
{\sc J.~Janssen et~al.}, {\em Numerical calculation of energy deposition by
  high-energy electron beams: {III}. three-dimensional heterogeneous media},
  Phys. Med. Biol., 39 (1994), pp.~1351--1366.

\bibitem{JanKorSto97}
\leavevmode\vrule height 2pt depth -1.6pt width 23pt, {\em Numerical
  calculation of energy deposition by high-energy electron beams: {III}-{B}.
  improvements to the 6d phase space evolution model}, Phys. Med. Biol., 42
  (1997), pp.~1441--1449.

\bibitem{Jet88}
{\sc D.~Jette}, {\em Electron dose calculation using multiple-scattering
  theory. {A}. {G}aussian multiple-scattering theory}, Med. Phys., 15 (1988).

\bibitem{JetBie89}
{\sc D.~Jette and A.~Bielajew}, {\em Electron dose calculation using
  multiple-scattering theory: Second-order multiple-scattering theory}, Med.
  Phys., 16 (1989).

\bibitem{KorAkhHei00}
{\sc E.~Korevaar et~al.}, {\em Accuracy of the phase space evolution dose
  calculation model for clinical 25 {M}ev electron beams}, Phys. Med. Biol., 45
  (2000), pp.~2931--2945.

\bibitem{Kri07}
{\sc H.~Krieger}, {\em Grundlagen der Strahlungsphysik und des
  Strahlenschutzes}, Teubner, 2.~ed., 2007.

\bibitem{LarMifFra97}
{\sc E.~Larsen et~al.}, {\em Electron dose calculations using the method of
  moments}, Med. Phys., 24 (1997).

\bibitem{LaVPim97}
{\sc J.~A. LaVerne and S.~M. Pimblott}, {\em Effect of elastic collisions on
  energy deposition by electrons in water}, J. Phys. Chem. A, 101 (1997),
  pp.~4504--4510.

\bibitem{LeaLar01}
{\sc C.~L. Leakeas and E.~W. Larsen}, {\em Generalized {F}okker-{P}lanck
  approximations of particle transport with highly forward-peaked scattering},
  Nucl. Sci. Eng., 137 (2001), pp.~236--250.

\bibitem{Lew50}
{\sc H.~W. Lewis}, {\em Multiple scattering in an infinite medium}, Phys. Rev.,
  78 (1950).

\bibitem{MarReyWag02}
{\sc C.~Martens et~al.}, {\em Underdosage of the upper-airway mucosa for small
  fields as used in intensity-modulated radiation therapy: A comparison between
  radiochromic film measurements, {M}onte {C}arlo simulations, and collapsed
  cone convolution calculations}, Med. Phys., 29 (2002), pp.~1528--1535.

\bibitem{MorHui92}
{\sc M.~Morawska-Kaczy\'nska and H.~Huizenga}, {\em Numerical calculation of
  energy deposition by broad high-energy electron beams: {II}. multi-layered
  geometry}, Phys. Med. Biol., 37 (1992), pp.~2103--2116.

\bibitem{Mor85}
{\sc J.~E. Morel}, {\em An improved {F}okker-{P}lanck angular differencing
  scheme}, Nucl. Sci. Eng., 89 (1985), pp.~131--136.

\bibitem{AAPMreport85}
{\sc N.~Papanikolau et~al.}, {\em Tissue inhomogeneity corrections for
  megavoltage photon beams}, Medical Physics Publishing, AAPM Report No. 85
  (2004).

\bibitem{Pom92}
{\sc G.~C. Pomraning}, {\em The {F}okker-{P}lanck operator as an asymptotic
  limit}, Mathematical Models and Methods in Applied Sciences, Vol. 2, No. 1
  (1992), pp.~21--36.

\bibitem{RosGre41}
{\sc B.~Rossi and K.~Greisen}, {\em Cosmic-ray theory}, Rev. Med. Phys., 13
  (1941), pp.~240--309.

\bibitem{Pen09}
{\sc F.~Salvat, J.~M. Fern\'andez-Varea, and J.~Sempau}, {\em PENELOPE-2008: A
  Code System for Monte Carlo Simulation of Electron and Photon Transport},
  OECD, 2009.

\bibitem{SchPfoJae06}
{\sc H.~Schwoerer et~al.}, {\em Laser-plasma acceleration of
  quasi-monoenergetic protons from microstructured targets}, Nature, 439
  (2006), pp.~445--448.

\bibitem{ShiHog91}
{\sc A.~S. Shiu and K.~R. Hogstrom}, {\em Pencil-beam redefinition algorithm
  for electron dose distributions}, Med. Phys., 18 (1991).

\bibitem{SiaWalDAl01}
{\sc C.~Siantar et~al.}, {\em Description and dosimetric verification of the
  peregrine {M}onte {C}arlo dose calculation system for photon beams incident
  on a water phantom}, Med. Phys., 28 (2001).

\bibitem{SpeLew08}
{\sc E.~Spezi and G.~Lewis}, {\em An overview of {M}onte {C}arlo treatment
  planning for radiotherapy}, Radiat. Prot. Dos., 131 (2008), pp.~123--129.

\bibitem{TerKolVau99}
{\sc J.~Tervo et~al.}, {\em A finite-element model of electron transport in
  radiation therapy and a related inverse problem}, Inverse Problems, 15
  (1999), pp.~1345--1361.

\bibitem{VasWarDav08}
{\sc O.~N. Vassiliev et~al.}, {\em Feasibility of a multigroup deterministic
  solution method for 3d radiotherapy dose calculations}, Int. J. Radiat.
  Oncol. Biol. Phys., 72 (2008), pp.~220--227.

\bibitem{VenWelMij01}
{\sc J.~Venselaar, H.~Welleweerd, and B.~Mijnheer}, {\em Tolerances for the
  accuracy of photon beam dose calculations of treatment planning systems},
  Radiother. Oncol., 60 (2001), pp.~191--201.

\bibitem{ZheBra93}
{\sc L.~Zheng-Ming and A.~Brahme}, {\em An overview of the transport theory of
  charged particles}, Radiat. Phys. Chem., 41 (1993), pp.~673--703.

\end{thebibliography}
\bibliographystyle{siam}

\end{document}